\documentclass[a4paper,11pt]{article}
\usepackage{pos}
\usepackage{booktabs}
\usepackage{aas_macros}

\title{Searching for joint neutrino and gravitational wave emission from the environment of Active Galactic Nuclei}
 \ShortTitle{Search for neutrino and gravitational wave emission from AGN}

 \author{Giacomo Bruno}
 \author{Gwenha\"el De Wasseige}
 \author{Romain Gorski}
 \author{Mathieu Lamoureux}
 \author*{Matthias Vereecken}

\affiliation{Centre for Cosmology, Particle Physics and Phenomenology - CP3,
Université Catholique de Louvain, Louvain-La-Neuve, Belgium}

\emailAdd{giacomo.bruno@uclouvain.be}
\emailAdd{gwenhael.dewasseige@uclouvain.be}
\emailAdd{romain.gorski@student.uclouvain.be}
\emailAdd{mathieu.lamoureux@uclouvain.be}
\emailAdd{matthias.vereecken@uclouvain.be}

\abstract{
With the observation of gravitational waves from merging compact
binary systems, a new observing window of the universe has been
opened. Most of the gravitational wave events currently detected are
due to the merger of binary black hole systems. One way to better
investigate such systems is to look for coincident emission in
electromagnetic waves or neutrinos. For typical models of isolated
binaries, no such emission is expected. However, one promising class
of mergers is that of binary black holes in the accretion disk of
active galactic nuclei. Such mergers potentially occur at high rates, since
these environments naturally have high numbers of black holes, which
can efficiently form binaries, merge rapidly, and potentially accrete
matter fast due to the surrounding gas. Here, we propose a method to
search for coincident gravitational wave and neutrino emission from
the location of known AGN, using an unbinned maximum likelihood
analysis, and apply it to currently available public data.
}

\ConferenceLogo{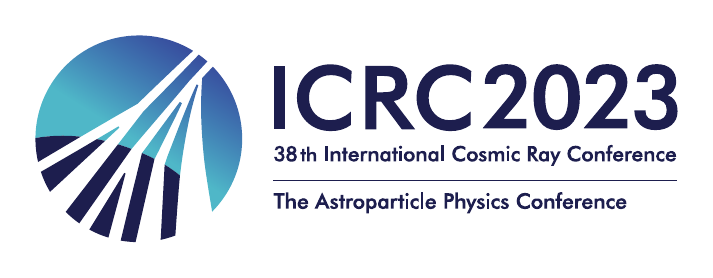}

\FullConference{%
38th International Cosmic Ray Conference (ICRC2023)\\
  26 July - 3 August, 2023\\
  Nagoya, Japan}

\begin{document}
\maketitle

\section{Introduction}

Gravitational waves were observed directly for the first time in 2015
by LIGO-Virgo-KAGRA (LVK). Since then, almost one hundred such
gravitational wave events have been detected. Most gravitational wave
events are the result of the merger of a binary black hole (BBH)
system, while a few of them are from a neutron star-black hole (NSBH)
or binary neutron star (BNS) system~\cite{LIGOScientific:2021djp}.
This dominance by BBH events, as well as their high masses, was unexpected.

For each of the gravitational wave events, there are also accompanying
searches for electromagnetic or neutrino emission by other
experiments.  For example, there exist several searches for coincident
emission of high-energy
neutrinos~\cite{IceCube:2022mma,ANTARES:2023wcj}.  Such a
(high-energy) multi-messenger signal is typically only expected for
mergers involving a neutron star, since they can supply the accreting
matter necessary for particle acceleration and subsequent interaction
and emission to occur.

However, tentative detections of possible coincident electromagnetic
(EM) signals have inspired models where such emission is possible for BBH
mergers~\cite{Perna:2016jqh,Graham:2020gwr}.  One class of such models
is of binary black holes merging in the accretion disk of active
galactic nuclei (AGN)~\cite{Bartos:2016dgn}. This region is expected
to contain many black holes, which can efficiently form binaries and
merge rapidly due to the surrounding gas. This also allows for
hierarchical mergers, where black holes undergo multiple mergers,
possibly explaining the high black-hole masses seen by LVK. Moreover,
the gas-rich environment allows for rapid accretion of gas and possible
emission of an electromagnetic and/or neutrino signal. Several
realizations of this scenario have been proposed to explain
multi-messenger emission, see
e.g.~\cite{McKernan:2019hqs,Graham:2020gwr}.
The rate at which BBH mergers in AGN accretion disks occur is
currently unknown, since there are many modeling
uncertainties. Estimates vary from less than a percent of the observed
BBH merger rate, to more than 50\% of the observed rate~\cite{Mckernan:2017ssq,Tagawa:2019osr,Grobner:2020drr,Ford:2021kcw}.




Establishing how frequently this scenario really occurs is
possible in several ways. Using localization alone, it is already
possible to show a coincidence between the location of BBH mergers and
AGN after less than a hundred BBH mergers at low
redshifts~\cite{Bartos:2017ggb,Veronesi:2022hql}. As of observing run
O3 of LVK, it can already be shown that the most luminous AGN most
likely do not produce the majority of BBH
mergers~\cite{Veronesi:2023ugk}. However, this does not significantly
constrain the total rate of BBH mergers in AGN.



In these proceedings, we aim to test whether BBH mergers occur in AGN
accretion disks and whether such mergers result in neutrino
emission. We propose a method, based on an unbinned maximum likelihood
analysis, to search for coincident emission of
gravitational waves and neutrinos from the location of known AGN.
In order to demonstrate our method, we apply it to currently available
public data from IceCube~\cite{IceCube:2019cia}, ANTARES~\cite{Aublin:2019zzn,ANTARES:2018osx},
and LVK~\cite{LIGOScientific:2018mvr}.

\section{Analysis}

\subsection{Data}

Since our analysis aims to find a correlation between gravitational
wave events, neutrinos, and active galactic nuclei, we need catalogs
for all three of these. For the neutrino data, we use the public
IceCube all-sky point-source dataset (2008--2018)~\cite{IcecubePointSource}
and
ANTARES point-source dataset
(2007--2017)~\cite{Aublin:2019zzn,ANTARES:2018osx}.
Both of these contain track events, which possess excellent pointing
as needed for a point-source search.
For the gravitational wave data, we use the gravitational wave transient
catalog (GWTC). Since the neutrino data we use only spans until
2018, we restrict ourselves to gravitational wave events in
GWTC-1~\cite{LIGOScientific:2018mvr}.
For every gravitational wave, LVK supplies a skymap that contains, for each pixel,
the probability $p\left(\Omega \left| \mathrm{GW} \right. \right)$ that the
gravitational wave comes from that direction and the probability
$p\left( d \left| \mathrm{GW} \right. \right)$ that the gravitational wave comes
from a source at distance $d$ if it comes from the direction $\Omega$. In
practice, the distance distribution is given by a Gaussian
parametrized by a mean and standard deviation $d_\mathrm{GW}\left(\Omega\right)$ and $\sigma_\mathrm{GW}\left(\Omega\right)$.

Finally, for the active galactic nuclei we test, we use the
V\'eron-Cetty AGN catalog~\cite{2010A&A...518A..10V}. While this
catalog does not contain data from the most recent surveys, it is
still a reliable source of AGN data. The catalog contains
168\,940~AGN, spread out over redshift, of which 95\,075 are at a
redshift smaller than 1.5, relevant for gravitational wave
coincidences. As with any AGN catalog, the biggest challenge is the
completeness of the catalog. In particular, some directions in the sky have been
surveyed far deeper than others. The effect of this is especially visible
at southern declinations, where the number of AGN in the catalog is small.

\subsection{Method}

In order to test our model, we propose a modified version of the unbinned
maximum likelihood method used in one of the combined gravitational
wave-neutrino searches by
IceCube~\cite{Hussain:2019xzb,IceCube:2022mma}.
At its core, we perform a neutrino point-source search at the location
of each AGN in our AGN catalog within the 90\% probability contour region
as inferred by LVK.
We consider neutrinos observed by IceCube and ANTARES within
a time window of 2~days centered on the observed arrival time of the
gravitational wave event.
To this neutrino point-source search, we add a prior based on the
inferred gravitational wave direction and distance.
Concretely, we construct the quantity
\begin{equation}
  \label{eq:postllh}
  \tilde{\mathcal{L}}^\mathrm{post}_\mathrm{AGN} = \mathcal{L}_{\nu, \mathrm{S+B}}\left( n_S \right) \times p\left(
    \Omega_\mathrm{AGN}\left| \mathrm{GW} \right. \right) \times p\left(
    d_\mathrm{AGN} \left| \mathrm{GW}  \right. \right),
\end{equation}
which can be interpreted as a sort of posterior likelihood.  Our method
differs from the one in~\cite{Hussain:2019xzb,IceCube:2022mma} through
the added distance factor and by testing individual AGN in a catalog
instead of performing a sky scan.

The first part of $\tilde{\mathcal{L}}^\mathrm{post}_\mathrm{AGN}$
represents the neutrino point-source likelihood. It is given by
\begin{equation}
  \label{eq:nullh}
  \mathcal{L}_{\nu, S+B}\left( n_S \right) = \prod_i^{N_\nu} \frac{n_S}{N_\nu}\mathcal{S}_i + \left(
    1 - \frac{n_S}{N_\nu} \right) \mathcal{B}_i,
\end{equation}
with $N_\nu$ the total number of neutrinos in the 2-day time window, $n_S$ the expected
number of signal neutrinos, $\mathcal{S}_i$ the signal probability distribution function,
and $\mathcal{B}_i$ the background probability distribution function.
For this first analysis, we only use a spatial likelihood. For each
observed neutrino in the time window, it consists of a signal term and a background term.
The signal likelihood gives the probability of measuring a neutrino direction
$\vec{x}_\mathrm{obs}$ from a true direction $\vec{x}_\mathrm{true}$,
and is described by the Von Mises-Fisher distribution,
which is the generalization of a Gaussian on a sphere. For
computational reasons, this distribution is approximated by a Gaussian
if the neutrino is well-localized (angular error below 7$^\circ$). The background distribution is
derived directly from the neutrino data and depends only on
declination.
This neutrino likelihood depends on $n_S$. Higher values of $n_S$ at
the tested position are preferred if the spatial distribution
of observed neutrinos prefers a point source at that position instead
of a distribution compatible with background.
Thus, for each AGN in our catalog within the gravitational wave localization
contour, we maximise the neutrino likelihood $\mathcal{L}_{\nu, S+B}\left( n_S
\right)$ in order to obtain the best-fit number of signal neutrinos
$n_S$ for that AGN.

The second factor in $\tilde{\mathcal{L}}^\mathrm{post}_\mathrm{AGN}$ is
$p\left(\Omega_\mathrm{AGN}\left| \mathrm{GW}  \right.\right)$, the
probability that the gravitational wave comes from the direction of
the AGN, as given by the gravitational-wave localization skymap.
Finally, the third factor in $\tilde{\mathcal{L}}^\mathrm{post}_\mathrm{AGN}$ is
$p\left(d_\mathrm{AGN} \left| \mathrm{GW}  \right. \right)$, the
probability that the gravitational wave source is located at the same distance as the
AGN, if it comes from the direction of the AGN. This probability is
given by
\begin{equation}
  \label{eq:pd}
  p\left(d_\mathrm{AGN} \left| \mathrm{GW}  \right. \right) =
  \frac{1}{\sqrt{2\pi}\sigma_\mathrm{GW}}\exp\left( -\frac{\left(
        d_\mathrm{AGN} - d_\mathrm{GW}\right)^2}{2\sigma_\mathrm{GW}^2} \right).
\end{equation}

For each gravitational-wave event, we then perform a hypothesis test of
our model against a background-only neutrino likelihood
(Eq.~\ref{eq:nullh} with $n_S=0$). We construct
a test statistic for each AGN compatible with the gravitational wave
event; it is given by
\begin{equation}
  \label{eq:TS}
  \Lambda_\mathrm{AGN} = 2\ln\left( \frac{\tilde{\mathcal{L}}^\mathrm{post}_\mathrm{AGN}}{\mathcal{L}_{\nu,\mathrm{B}}} \right).
\end{equation}
The AGN with the highest value of the test statistic
$\Lambda_\mathrm{AGN}$ is considered as the best-fit AGN for that particular
gravitational-wave event.

In order to obtain the significance of the best-fit result, we repeat the
analysis above for 10\,000 trials with randomized neutrino
skymaps. These randomized neutrino skymaps are obtained by scrambling
the neutrinos in the IceCube and ANTARES catalogs in time, and
reselecting the neutrinos in the time window of the gravitational wave
event. In this way, we can estimate the $p$-value, i.e.\ the chance
that a background-only neutrino skymap produces a best-fit result with
a test statistic equal to or higher than the one observed.

\section{Results}

Our results are shown in Table~\ref{tab:results}. None of the
gravitational wave events in GWTC-1 show a significant excess of
neutrinos at the location of an AGN in the V\'eron-Cetty catalog.  The
completeness of the AGN catalog strongly influences the
results. Indeed, when a gravitational-wave localization contour only
contains a few AGN in the catalog, the best-fit number of neutrinos is
often equal to zero. The only exception is GW170818, which only has
one coincident AGN but has the lowest $p$-value. However, since the
$p$-value is still high and this region is poorly surveyed by the AGN
catalog, we do not consider this result as significant. Likewise, the
gravitational-wave localization also has a strong effect on our
analysis, since large localisations imply a high number of AGN to test
which increases the chance of an accidental coincidence with a
clustering of background neutrinos, which in turn decreases the
sensitivity of the analysis.

\begin{table}
  \centering
  \begin{tabular}{lrrrlrr}
\toprule
GW event & Area (deg$^{2}$)& $N_\nu$ & $N_\mathrm{AGN}$ & best-fit AGN & $n_S$ & $p$-value \\
\midrule
GW150914 & 182 & 785 & 2 & IRAS 03230-5800 & 0.00 & 1.00 \\
GW151012 & 1523 & 733 & 257 & MARK 1187 & 1.31 & 0.96 \\
GW151226 & 1033 & 767 & 282 & SDSS J12548-0010 & 0.90 & 0.63 \\
GW170104 & 921 & 745 & 1096 & 2MASS J08191+3419 & 1.00 & 0.75 \\
GW170608 & 392 & 684 & 213 & 3C 192 & 0.00 & 1.00 \\
GW170729 & 1041 & 722 & 151 & SDSS J12254+4901 & 1.40 & 0.27 \\
GW170809 & 308 & 710 & 58 & Q 0052-2956 & 0.00 & 1.00 \\
GW170814 & 87 & 699 & 7 & RXS J03149-4241 & 0.00 & 1.00 \\
GW170818 & 39 & 707 & 1 & MARK 308 & 0.53 & 0.20 \\
GW170823 & 1666 & 692 & 357 & MS 07199+7100 & 1.60 & 0.47 \\
\bottomrule
\end{tabular}
  \caption{Results of our analysis, showing for each gravitational
    wave event its localization, the number of neutrinos in a 2-day
    time window, the number of AGN in the V\'eron-Cetty catalog within
    the 90\% contour region, the best-fit AGN, the number of fitted signal
    neutrinos, and the significance (p-value) of the best fit.}
  \label{tab:results}
\end{table}

\section{Discussion}

The current analysis serves as a demonstration of the method we
propose to search for a coincidence between neutrinos, gravitational
waves, and AGN. The three ingredients to our analysis each add their
own information: gravitational waves contain some distance information
but are badly localized, neutrinos have no distance information but
improved pointing, while AGN have accurate position and distance. In
this way, this method can test whether binary black hole mergers occur
regularly in
AGN accretion disks and whether this results in neutrino emission.

However, the current analysis can be improved in several ways. First,
in order to demonstrate the analysis, we only included spatial
information in the neutrino likelihood. A full analysis
should also make use of the energy distribution and the number of
observed neutrinos compared to the expected background rate.

More critically, the incompleteness of the AGN catalog needs to be
addressed (see also~\cite{Veronesi:2023ugk}). This aspect can be improved in
multiple ways: using more recent/complete surveys, combining different AGN catalogs,
testing only AGN with high luminosity (assuming that they also have more
binary black hole mergers and/or gas powering neutrino emission, and
more likely to be complete in catalogs), and considering
gravitational wave events only up to a certain redshift, with good
localization, or in regions of the sky where catalogs are sufficiently complete.

\acknowledgments
M.L.\ is a postdoctoral researcher of the Fonds de la Recherche
Scientifique - FNRS. G.D.W. acknowledges the support from the
Francqui Foundation.

\bibliographystyle{JHEP}
\bibliography{biblio.bib}



%
%
%

\end{document}